\documentclass[aps,prd,groupedaddress,11pt,showpacs]{revtex4}
\usepackage{graphicx,epsf,amssymb,amsmath,latexsym}
\newcommand{\p}{/\!\!\!\! p}
\newcommand{\nb}{\bar/\!\!\!\!\nabla}

\newcommand{\D}{/\!\!\!\! D}
\newcommand{\be}{\begin{eqnarray}}
\newcommand{\ee}{\end{eqnarray}}
\renewcommand{\d}{\partial}

\numberwithin{equation}{section}

\begin{document}

\begin{flushright}
DAMTP-2004-140
\end{flushright}

%\vspace*{.5in}
%
\title{Higher-Spin Fields in Braneworlds}

\vspace{.3in}

\author{Cristiano Germani}
\email{C.Germani@damtp.cam.ac.uk}
\affiliation{D.A.M.T.P., Centre for Mathematical Sciences, University of Cambridge\\
Wilberforce road, Cambridge CB3 0WA, England }
\author{Alex Kehagias}
\email{kehagias@central.ntua.gr}
\affiliation{Physics Division, National Technical University of Athens, \\
15780 Zografou Campus,  Athens, Greece }
\vskip.3in
\begin{abstract}
The dynamics of higher-spin fields in braneworlds is discussed.
In particular, we study fermionic and bosonic
higher-spin fields in $AdS_5$ and their localization
on  branes. We find that  four-dimensional zero modes exist only for
spin-one fields, if there are no couplings to the boundaries.
If boundary couplings are allowed, as in the
case of the bulk graviton,  all bosons acquire a  zero mode
irrespective of their spin. We  show that there are boundary
conditions for fermions, which generate chiral zero modes in the
four-dimensional spectrum. We also propose a gauge
invariant on-shell
action with cubic interactions by adding non-minimal couplings, which depend on the Weyl tensor.
In addition, consistent couplings between higher-spin fields and matter on the brane are presented.
Finally, in the $AdS$/$CFT$ correspondence, where bulk 5D theories on $AdS$ are related to 4D $CFT$s,
we explicitly discuss the holographic picture of higher-spin theories
in $AdS_5$ with and without boundaries.
\end{abstract}
\maketitle
%
%\newpage

\section{Introduction}

The problem of consistent higher-spin (HS) gauge theories is a
fundamental problem in field theory. After the description of
their free dynamics~\cite{Fronsdal:1978rb}, \cite{Fang:1978wz},
only negative results for their interactions were
obtained~\cite{Aragone:1979hx},\cite{NWE}. For example, it was realized
 that HS fields cannot consistently minimally interact with gravity.
However, by allowing additional gaugings, one may introduce
counter terms, which make the propagation of HS fields
in curved backgrounds well-defined. By appropriate completion of the interactions,
Vasiliev equations are found, which are the generally covariant field equations for massless
HS gauge fields describing their consistent interaction with gravity \cite{Vass},\cite{Vasss},\cite{Sezgin}.

 Nowadays, there is
a renewal interest in HS gauge theories.  A basic reason
for this is that HS theories
 exist on
anti-de Sitter spaces $AdS$ \cite{Fronsdal:1978vb}, signaling
their relevance in the $AdS$/$CFT$ correspondence. In this
framework, as a general rule, conserved currents in the boundary
$CFT$ are expected to correspond to massless gauge fields in the
bulk \cite{ref}. A weakly coupled boundary gauge theory for example contains
an infinite number of almost conserved currents, which will be
described by a dual HS gauge theory defined in the bulk of $AdS$.
Although much remain to be done in this direction, specific
progress has been made in three-dimensional $CFT$s.  It was
proposed in~\cite{Polyakov-Klebanov} for example, that the singlet
sector of the three-dimensional critical $O(N)$ vector model is
dual, in the large $N$ limit, to a minimal theory in
four-dimensional anti-de Sitter space containing massless gauge
fields of even spin of the kind studied in~\cite{Vass}. String
theory also gives additional support to HS fields.
Indeed, string theory,   contains an infinite number of massive
HS fields with consistent interactions. In the
low-tension limit, their masses disappear. Massless HS
theories are thus the natural candidates for the description of
the low-tension limit of string theory at the semi-classical
level~\cite{Ulf}. The hope is that the understanding of the
dynamics of HS fields could help towards a deeper insight
of string theory, which now is mainly based on its low-spin
excitations and their low-energy interactions.

A generic massless bosonic particle of integer spin $s$ in an
n-dimensional spacetime is described by a  totally symmetric
tensor of rank $s$, $\phi_{\mu_1\mu_2...\mu_s}$, while a fermionic
particle of spin $s+\frac{1}{2}$  by a totally
symmetric tensor-spinor of rank $s$, $\psi_{\mu_1\mu_2...\mu_s}$.
These fields are defined up to gauge transformations and they are subject to certain
constraints such that the corresponding theories are ghost free.
This means that they describe exactly two propagating
modes of $\pm s$ and $\pm (s+\frac{1}{2})$ helicities, for bosons
and fermions, respectively. Such theories may be obtained as the
massless limits~\cite{Fronsdal:1978rb},\cite{Fang:1978wz} of
massive HS theories~\cite{Singh:1974qz} or by gauge invariance and
supersymmetry, as the latter relates HS fields to known
lower spin ones~\cite{Curtright:1979uz}.

In flat Minkowski spacetime, the gauge transformations of the HS fields are
\be
&&\delta \phi_{\mu_1\mu_2...\mu_s}=
\partial_{(\mu_1}\xi_{\mu_2\mu_3...\mu_s)}\, ,~~~~~~~
%\\ &&
\delta \psi_{\mu_1\mu_2...\mu_s}=
\partial_{(\mu_1}\epsilon_{\mu_2\mu_3...\mu_s)}\, ,
\ee
where the
parenthesis denote the symmetrized sum of s-terms (without the
usual combinatorial  $s!$) and $\xi_{\mu_2\mu_3...\mu_s}, \,
\epsilon_{\mu_2\mu_3...\mu_s}$ are totally symmetric rank-$(s-1)$
tensor and tensor-spinors, respectively. In addition, we impose on
these fields the strongest gauge invariant constraints
\be
&&{\phi^{\mu\nu}}_{\mu\nu\mu_5...\mu_s}=0\, ,~~~~~~~
%\\&&
\gamma^\nu{{\psi_{\nu}}^\mu}_{\mu\mu_4...\mu_s}=0\, , \label{dt}
\ee
which means that the
 bosonic HS fields are  double traceless,
while the fermionic ones are triple $\gamma$-traceless (as a trace in the fermionic conditions
can be considered as due to two $\gamma$ matrices). These conditions  give
constraints for $s\geq 4$ and $s\geq \frac{7}{2}$ for bosons and
fermions, respectively and
eliminate their lower-spin components. In addition,
one can impose traceless and $\gamma$-traceless of the
gauge parameters $\xi,\epsilon$, respectively, i.e.,
\be
&&{\xi^\nu}_{\nu\mu_4...\mu_{s}}=0\, , ~~~~~~
%\\&&
\gamma^\nu{{\epsilon_{\nu}}}_{\mu_3\mu_4...\mu_s}=0\, .
\ee
It should be noted, however,
that there exists also a
recently proposed formulation~\cite{Francia:2002aa},\cite{Sagnotti:2003qa},
where the gauge parameters are not constrained.

A simple counting
reveals that there are only two independent degrees of freedom for both the bosonic and fermionic
HS fields. In this case, consistent ghost-free equations of
motions for the massless gauge HS fields can be written
down, which described the propagation of the two helicity modes of these fields~\cite{deWit:1979pe},\cite{Sorokin}.
In particular, pure gauge degrees of freedom can be eliminated by imposing, for integer
spins, the  appropriate generalizations of the
Lorentz and de Donder gauges
\be
&&\d_\nu {\phi^\nu}_{\mu_2\mu_3...\mu_s}-\frac{1}{2} \d_{(\mu_2}{\phi^\nu}_{\nu\mu_3...\mu_s)}=0\, ,\label{gaugeb}
\ee
whereas the corresponding gauge conditions for half-integer fields reads
\be
&&\gamma^\nu {\psi}_{\nu\mu_2\mu_3...\mu_s}-\frac{1}{2s} \gamma_{(\mu_2}{\psi^\nu}_{\nu\mu_3...\mu_s)}=0\, .
\label{gaugec}
\ee
In this case, the bosonic $\phi_{\mu_1\mu_2...\mu_s}$ and the
fermionic $\psi_{\mu_1\mu_2...\mu_s}$ fields satisfy
\be
&&\Box \phi_{\mu_1\mu_2...\mu_s}=0\, ,~~~~~~~~/\!\!\!\d \psi_{\mu_1\mu_2...\mu_s}=0\, .
\ee
Thus, $ \phi_{\mu_1\mu_2...\mu_s}, \, \psi_{\mu_1\mu_2...\mu_s}$ indeed describe massless particles, as claimed.

It is clear from the above that there is no problem of writing down HS field
equations in flat space for free fields. The problems appear when one
considers interactions of these fields. The most obvious
interaction is the gravitational interaction.  An immediately way
of introducing the latter  is to replace ordinary derivatives with
covariant ones in order to maintain general covariance. However,
in this case gauge invariance is lost as we need to commute
derivatives in the field equations~\cite{Aragone:1979hx}.
In fact only in
flat Minkowski spacetime derivatives commute and gauge invariance is possible.
Indeed, for massless fields of spin $s>\frac{1}{2}$, the field
equations for bosons take schematically the form $D_\mu F^{\mu\mu_2..}=0$, where
$F^{\mu\mu_2...}$ is the antisymmetric field strength, a
generalization of the Maxwell $F^{\mu\nu}$ tensor~\cite{deWit:1979pe}. Then the
Bianchi identity $D_\mu D_\nu F^{\mu\nu\mu_3...}=0 $ leads, by the
non-commutativity of the covariant derivatives, to local
constraints of the form $W_{\mu\nu\kappa\lambda}F^{\mu\nu...}=0$.
As these constraints
involve the Weyl tensor, i.e, the part of the  Riemann tensor
which is not specified by the gravitational field equations, minimal coupling
of such field to gravity are not in general consistent. An exception is for
spin $s=1,2$, which involve only Ricci and curvature scalar terms.
The same happens with half-integer HS fields. This means
that HS fields minimally couplet to gravity have acausal propagation in curved
spacetimes and cannot consistently be defined. As a general rule, gauge invariance and general
covariance cannot be simultaneously imposed, indicating
the inconsistency of a minimal couplings of HS fields with spin $s>2$  to
gravity. This "no-go theorem'' can however be circumvented
on backgrounds with
vanishing Weyl tensor, i.e., on conformally flat space-times, such
as de Sitter ($dS$) and anti-de Sitter ($AdS$) spacetimes~\cite{Deser}. Indeed,
soon after the results of
\cite{Fronsdal:1978rb},\cite{Fang:1978wz}, propagation of
HS fields on (A)dS have been discussed
in~\cite{Fronsdal:1978vb}.
In particular, by  gauging  an infinite-dimensional generalization of the target
space Lorentz algebra, consistent interactions
of HS fields has been introduced~\cite{Vasss},\cite{Sezgin}. However such consistent interactions do not have
a flat spacetime limit as they are based on generally covariant curvature expansion on $(A)dS$ spacetime with
expansion parameter proportional to the $(A)dS$ length.

Here we will discuss HS fields living not in the whole of
$AdS$ spacetime, but rather in a part of it.
A particular example of such spacetimes, once boundaries are
introduced, is the Randall-Sundrum one~\cite{Randall:1999ee},
which has extensively be studied as an alternative to
compactification and in connection with the hierarchy
problem~\cite{AK}.
The aim is twofold. Firstly, to study localization
properties of HS fields in the 4D boundary of the anti-de
Sitter space and secondly, to relate bulk fields to HS
operators in the dual boundary theory.
For this, we will study the reduced 4D theory
for HS fields. In this case, HS fields may also have consistent gravitational
interactions on flat spacetime (brane), although the whole tower
of massive KK modes of the bulk fields are needed. Moreover, we
find  cubic couplings of the HS fields to gravity by
introducing non-minimal terms in the action, as well as possible
couplings of the HS fields to matter living in the brane.

In the RS2 background the holographic boundary theory is a strongly coupled $CFT$ defined with a cutoff
and coupled to 4D gravity, whereas in
 RS1, the boundary theory is a badly broken $CFT$ in the IR~\cite{RZ}. In this framework, we will
examine the holographic description in RS backgrounds as well as in the $AdS$ spacetime.

 In the next section 2, we discuss briefly the
geometric setup and the boundary conditions needed. In section 3 and 4, we solve
the HS bulk equations and we find the 4D spectrum
for bosons and fermions, respectively. In sections 5, we present a gauge invariant on-shell action with cubic interactions
and couplings of the HS fields to matter on the brane. In section 6, we discuss the holography in $AdS_5$
for HS bosons and in
RS for HS fermions. Finally, in section 7, we conclude by summarizing our results.

\section{Higher spins in a box: $ AdS_5$ with boundaries}

We will mainly consider here five-dimensional anti-de Sitter
bulk spacetime $AdS_5$ with four-dimensional boundaries.
In this case and in order to set up the notation,
let us recall that $AdS$ is a maximally symmetric spacetime. Its
Riemann tensor is given in terms of its metric as
\be
R_{\alpha\beta\mu\nu}=-\frac{\Lambda}{4}(g_{\alpha\mu}g_{\beta\nu}-g_{\alpha\nu}g_{\beta\mu})
\, \label{rmn}
\ee
where $-\Lambda<0$
is the five-dimensional cosmological constant.
In  Gaussian-normal coordinates, the metric takes the form
\be
ds^2=e^{-2\sigma}\eta_{ab}dx^a dx^b+dy^2\ , \label{metric}
\ee
where $a,b,...=0,...,3$, $y=x_5$ and $\sigma=\sigma(y)$. In this coordinates, eq.(\ref{rmn}) gives
\be
\sigma'=\pm\sqrt{\Lambda}/2\, , ~~~\sigma''=0\, , \label{s0}
\ee
where  a prime
denotes  derivative with respect to the normal
coordinate $y$, i.e.,$ (')=\d/\d y$. For a smooth $AdS_5$ spacetime, eq.(\ref{s0})
is solved for $\sigma=\sqrt{\Lambda}/2 \, y$ with
$-\infty<y<\infty$.
A Randall-Sundrum (RS) background~\cite{Randall:1999ee} now,  is an
 $AdS$ spacetime invariant under $y\to -y$.
As a result, it may be viewed as a $Z_2$ orbifolds of $AdS$, and thus,
only  its positive section $y\geq 0$  may  be considered. In addition, in the first model
(RS1) of~\cite{Randall:1999ee},
there exists an ``end of the
world"  at $y=\pi R$ so that $0\leq y\leq \pi R$. By taking the limit
 $R\rightarrow\infty$, the second boundary at $y=\pi R$ is removed to infinity and one ends up with the second
 model (RS2) of~\cite{Randall:1999ee}.
Hence,
in the RS1 background we have $0\leq y\leq \pi R$, while
$0\leq y< \infty$ in RS2. On the other hand,  for both RS1 and RS2 models, the positive root of
$\sigma'=\sqrt{\Lambda}/2$ for $y>0$ and the negative one for $y<0$ may be used, so that we have
$$\sigma=2 a |y|\, , ~~~~ a=\sqrt{\Lambda}/4.$$
However, with this form of the wrap factor, the second derivative of $\sigma$, which enters in the curvature tensors
 does not vanish but rather gives
 $\delta$-function contributions to both Riemann and Ricci tensors. These contributions may be cancelled
by putting branes of appropriate fine-tuned tensions at the fixed points of the $Z_2$ orbifold. The branes are 4D flat
Minkowski spacetimes and they are the boundaries of the
bulk $AdS$ background.
The boundary at $y=0$ is the
UV brane whereas the brane at $y=\pi R$ is the IR one.  In RS1 our Universe is on the IR brane. In this
way a possible solution of the hierarchy problem has been
suggested. However, here a negative Newtonian constant appear. In
the second model RS2 of~\cite{Randall:1999ee} instead (where $R\rightarrow\infty$),  our
Universe is on the UV brane as the IR one has been removed to infinity. This model is considered as a valid
alternative to compactification.

The fields living in the bulk are as if they were propagating in $AdS$ but now they
will, in addition,  experience two boundaries at $y=0$ and
$y=\pi R$ in RS1, or just one boundary in RS2.
Our task is to discuss the localization problem and the effective theory living
on these 4D boundaries for integer spin fields with
spin $s\geq 1$ and semi-integer spinors with spin $s\geq 3/2$.
Spin $0,1$ and $1/2,3/2$ are particular cases and have already been
discussed~\cite{Pomarol:1999ad},\cite{9912232}.

In curved spacetime, one has to modify the definition of the spacetime covariant derivative
in order to maintain a local Lorentz invariance
of a semi-integer spin field. This is achieved by introducing the covariant
derivative
\be D_\mu=\nabla_\mu+\Gamma_\mu\ , \ee
where $\nabla_\mu$ is the spacetime covariant derivative and the spin
connection is defined as
\be
\Gamma_\mu=\frac{1}{2}\Sigma^{ab}e^\nu_a\nabla_\mu e_{b\nu}\ .
\ee
Here $\Sigma^{ab}$ are the local generators of Lorentz symmetry and
$e_a^\mu$ is the n-bein.
For an  $AdS_{p+1}$ spacetimes with cosmological constant $-\Lambda\, , (\Lambda>0)$,
one may introduce the $SO(2,p)$-covariant derivative
\be \bar
\nabla_M=D_M+\left(\frac{\Lambda}{4 p}\right)^{1/2}\gamma_M\ , ~~~~M=0,...p\ , \ee
where $\gamma_M$ are the (p+1)-dimensional gamma  matrices.
In particular, the $SO(2,4)$-covariant derivative for $AdS_5$ is
\be \bar
\nabla_\mu=D_\mu+a\gamma_\mu\ ,\ee
and, in Gaussian-normal coordinates, the spin connection in $AdS_5$ is
\be
\Gamma_a=\frac{1}{2}\gamma_5\gamma_a\ \sigma' ,~~~~~ \Gamma_5=0\ .
\ee
Defining as usual $ \gamma^{\mu\nu}=\frac{1}{2}[\gamma^\mu,\gamma^\nu]$
and using the relation $[\gamma_\nu,D_\nu]=0$,
 a straightforward calculation  explicitly
shows that for a fermion $\Psi$
\be
[D_\mu,D_\nu]\Psi=\frac{1}{4}R_{\mu\nu\alpha\beta}\gamma^{\alpha\beta}\Psi\ .
\ee
For a general tensor-spinor $\epsilon_{\alpha_1...\alpha_s}$ of rank s, the $SO(2,4)$-covariant derivative
 satisfies
\be [\bar
\nabla_\mu,\bar\nabla_\nu]\epsilon_{\alpha_1...\alpha_s}&=&[D_\mu,D_\nu]\epsilon_{\alpha_1...\alpha_s}+2a^2\gamma_{\mu\nu}
\epsilon_{\alpha_1...\alpha_s}\, .
\ee

A central issue when boundaries are present, as in the RS background,  is the boundary condition problem.
The variation of the action introduces
boundary terms, which appear in
the integration by parts during the variational process.
These boundary terms must vanish independently from the bulk
terms, which provide the equations of motion, and introduce appropriate boundary conditions. For fermionic fields for example,
the action is of the form
\be
S= \frac{1}{2}\int d^Dx\sqrt{-g}\Psi_{\alpha_1...\alpha_{s-1/2}}\gamma^\beta\bar\nabla_{\beta}\Psi^{\alpha_1...\alpha_{s-1/2}}+...\ .
\ee
In the presence of boundaries, the variation of the above action provides  boundary terms, which should
vanish
\begin{eqnarray}
\left(\delta \Psi^L\cdot\Psi^R-\delta \Psi^R\cdot
\Psi^L\right)\Big|_{0,\pi R}=0.
\end{eqnarray}
We have denoted by a dot the inner product, i.e., ($\Psi\cdot \Psi=\Psi_{\alpha_1...\alpha_{s-1/2}}\Psi^{\alpha_1...\alpha_{s-1/2}}$).
As we are interested in the  $Z_2$ symmetry $y\to -y$, it is
easy to see that the action $S$ is $Z_2$ symmetric if $\Psi(-y)=\pm \gamma^5 \Psi(y)$, or
\begin{eqnarray}
\Psi^L(-y)=\pm \Psi^L(y)\ ,\cr
\Psi^R(-y)=\mp \Psi^R(y)\ .
\end{eqnarray}
Without lost of generality we can choose the positive root. This means that the right-handed
field will in general have a ``kink'' profile
passing throughout $y=0$. Considering therefore only the positive domain $y>0$, one can use the following boundary
conditions
\be
\begin{array}{rl}
\mbox{(i)}~&\Psi^L(\pi R)=\Psi^R(\pi R)\, , ~~~
\Psi^L(0^+)=\Psi^R(0^+)\, , \\ \nonumber
\mbox{(ii)}~&\Psi^L(0^+)=\Psi^L(\pi R)=0\, ,\\ \nonumber
 \mbox{(iii)}~&\Psi^R(0^+)=\Psi^R(\pi R)=0\, , \label{sbc}
 \end{array}
\ee
However the boundary conditions (ii) and (iii) can be modified allowing a non-zero mass at the UV boundary.
This mass term at the boundary will be crucial for the holographic
interpretation as we shall see later.

With a similar procedure, we can consider very schematically a bosonic field with action
\be
S=\frac{1}{2} \int d^Dx\sqrt{-g}\nabla_{ \mu}\Phi_{ \alpha_1...\alpha_{s}}\nabla^{\mu}\Phi^{ \alpha_1...\alpha_{s} }+...\ .
\ee
Without the $Z_2$ symmetry, the variational principle, in gaussian-normal coordinates,
 is well defined if
\be
\left(\delta \Phi\cdot n^a\d_a\Phi\right)\Big|_{0,\pi R}=0\ .
\ee
However as the spacetime is $Z_2$ symmetric, the bulk field variation has a term like
\be
\delta \Phi_{\alpha_1...\alpha_{s}}\nabla_\mu\nabla^{\mu}\Phi^{ \alpha_1...\alpha_{s}}\ ,
\ee
which in fact contain a boundary term on the fix points of the spacetime.
This happens because the second derivative of the metric is distributional on the fix points. Such
second derivative is coming from terms containing $n^a\d_a\Gamma $, where
we schematically use $\Gamma$ to mean the discontinuous part of
the Christoffel symbols. With that a boundary term like
\be
(s-1)\delta \Phi\cdot\Phi \, n^a\d_a \Gamma \ ,  \label{ghg}
\ee
arises.
Then, in gaussian-normal coordinate on an $AdS$ background,
we obtain the following two possible boundary conditions for a bosonic field $\Phi$ of any spin
\vskip.1in
\paragraph{\rm Neumann}
\be
\Phi'(y)+(s-1)\, \sigma'\Phi(y)\Big|_{0,\pi R}=0\ , \label{NN}
\ee

\paragraph{\rm Dirichlet}

\be
\Phi(y)\Big|_{0,\pi R}=0\ . \label{DD}
\ee

\section{Integer spins in braneworld}

A generic field of spin s is described by a totally symmetric
rank-s tensor $\phi_{\mu_1\mu_2...\mu_s}$. As we shall discuss
briefly at the end of this section and in Section $6$, its field
equations on a smooth $AdS_{p+1}$ spacetime may be written as
\be
\nabla^2\phi_{\mu_1\mu_2...\mu_s}-M^2\phi_{\mu_1\mu_2...\mu_s}=0\, ,
\label{sin}
\ee
where the covariant derivatives are with respect to the $AdS$ background.
It can in fact be proven that eq.(\ref{sin})
is invariant under the gauge transformation
\be \label{df}
&&\delta \phi_{\mu_1\mu_2...\mu\_s}=\nabla_{(\mu_1}\xi_{\mu_2\mu_3...\mu_s)}
\ee
only for the particular value
$$M^2=\frac{\Lambda}{p}\left(s^2\!-\!(5\!-\!p) s\!-\!2p\!+\!4\right), $$
provided  $\phi_{\mu_1\mu_2...\mu_s}$ satisfies the gauge condition
\be
\nabla_\nu {\phi^\nu}_{\mu_2\mu_3...\mu_s}-\frac{1}{2}
\nabla_{(\mu_2}{\phi^\nu}_{\mu_3...\mu_s)\nu}
=0\, . \label{ds11}
\ee
This may easily be verified by taking into account that
\be
\nabla^2\xi_{\mu_2\mu_3...\mu_s}=\frac{\Lambda}{p}(p+s-2)(s-1) \xi_{\mu_2\mu_3...\mu_s} \, ,
\ee
which follows from eqs.(\ref{df},\ref{ds11}). In particular, for
$AdS_5$ ($p=4$), gauge invariance is
achieved for
\be
M^2=4 a^2 (s^2-s-4)\, .
\ee

However, in a RS background, the HS field equations eq.(\ref{sin})
are not invariant under the gauge transformation eq.(\ref{df}).
The reason is that in this case there are $\delta$-function
contributions coming from
 the Riemann and Ricci tensors. These contributions spoil gauge invariance, which can be restored,
nevertheless, by adding appropriate terms to the field equation
(\ref{sin}). For example,  the gauge variation of (\ref{sin}) for the spin s component of the
reduced field $\phi_{m_1...m_s}$, (Latin indices take the values
$m,n=0,...3$), turns out to be
\be
\delta(\nabla^2\phi_{m_1...m_s}-M^2\phi_{m_1...m_s})=-4a (s-2)\delta(y)
\delta\phi_{m_1...m_s} \, ,
\label{sin1}
\ee
so that the gauge invariant field equations for HS fields in RS background for $\phi_{m_1...m_s}$ is
\be
\nabla^2\phi_{m_1...m_s}-M^2\phi_{m_1...m_s}+4a (s-2)\delta(y)
\phi_{m_1...m_s}=0 \, .
\label{sin2}
\ee
It turns out, after explicitly calculating (\ref{sin2}), that the above field equations,
 in the gauge $\phi_{5\mu_2...\mu_s}=0$, are  written as
\be e^{2 \sigma}\d_m\d^m
\phi_{m_1...m_s}\!+\!\phi''_{m_1...m_s}\!+\!2(2\!-\!s)\sigma'
\phi_{m_1...m_s}'\!+\!\left((s(s\!-\!1)\!-\!4s)\sigma'^2\!-\!M^2\right)\phi_{m_1...m_s}=0
\ee supplemented with the boundary condition
\be
\phi'_{m_1...m_s}+4\, a\,  (s\!-\!1)\,
\phi_{m_1...m_s}\Big{|}_{0,\pi R}=0\, . \label{bc0}
\ee
  We may write, denoting collectively indices by dots,
\be
\phi_{...}(x,y)=\sum_n f_n(y) \varphi^{(n)}_{...}(x)
\ee
where $\varphi_{...}(x)$ is an ordinary massive spin-s field in Minkowski spacetime
\be
\d_m\d^m \varphi^{(n)}_{...}(x)=m_n^2\varphi^{(n)}_{...}(x)\, .
\ee
Then $f_n(y)$ satisfies the equation
\be
e^{2\sigma} m_n^2f_n \!+\!f_n''\!-\!4a(2\!-\!s) f_n'\!+\!16a^2(1\!-\!s)f_n=0\, , \label{eq1}
\ee
with  boundary conditions
\be
f_n'+4 a  (s\!-\!1)\, {f_n}|_{0,\pi R}=0\, . \label{bc}
\ee
The solution of eq.(\ref{eq1}) is
\be
f_n\!=\!\frac{e^{2a(2\!-\!s)|y|}}{N_n}\!\left( \!J_s\Big{(}\frac{m_n\,e^{2a |y|}}{2a}\Big{)}\!+
\!b_\nu(m_n) Y_s \Big{(}
\frac{m_n\,e^{2a |y|}}{2a}\Big{)}\!\right)\, , \label{fn}
\ee
where $\nu$ is the order of the Bessel's functions appearing in the solution.

For completeness, it should be noted that the  corresponding solution in a $(p+1)$-dimensional space
 $AdS_{p+1}$ is
\be
f_n\!=\!\frac{e^{a (p \!-\! 2 s) |y|}}{N_n}\!\left( \!J_{2-s-\frac{p}{2}}
\Big{(}\frac{m_n\,e^{2a |y|}}{2a}\Big{)}\!+
\!b_\nu(m_n) Y_{2-s-\frac{p}{2}} \Big{(}
\frac{m_n\,e^{2a |y|}}{2a}\Big{)}\!\right)\, , \label{ge}
\ee
which clearly reduces to eq.(\ref{fn}) for $p=4$.

For canonically normalized 4D fields $\varphi^{(n)}_{...}$, the normalization of
 $f_n$ in RS1 should be
\be
\int_{0}^{\pi R}dy\,  e^{4a(s-1)y} f_n^*f_m=\delta_{mn}\, .
\ee
Therefore, the parameter $N_n$ in eqs.(\ref{fn}) are
\be
N_n^2=
\int_0^{\pi R}\!\!\!\!\!dy\, e^{2\sigma}\!\!\left[ \!J_s\!\Big{(}\!\frac{m_n\,e^{2a |y|}}{2a}\!\Big{)}\!\!+\!
\!b_\nu Y_s \!\Big{(}\!
\frac{m_n\,e^{2a |y|}}{2a}\!\Big{)}\!\right]^2\, ,
\ee
and by employing the boundary conditions (\ref{bc})
we get the relations
\be
&&b_\nu(m_n)=-\frac{s J_\nu(\frac{m_n}{2a})+
\frac{m_n}{2a}J_\nu'(\frac{m_n}{2a})}{s Y_\nu(\frac{m_n}{2a})+\frac{m_n}{2a}Y_\nu'(\frac{m_n}{2a})}\, ,
\nonumber \\ &&
b_\nu(m_n)=b_\nu(m_ne^{2a \pi R}) \,  . \label{bc1}
\ee
Accordingly, for the $(p+1)$-dimensional space
 $AdS_{p+1}$, applying (\ref{bc}) to the general solution (\ref{ge}), we get
 \be
&&b_\nu(m_n)=-\frac{(2s-p+4) J_\nu(\frac{m_n}{2a})+
\frac{m_n}{a}J_\nu'(\frac{m_n}{2a})}{(2s-p+4) Y_\nu(\frac{m_n}{2a})+\frac{m_n}{a}Y_\nu'(\frac{m_n}{2a})}\, , \nonumber
\\ &&
b_\nu(m_n)=b_\nu(m_ne^{2a \pi R}) \,  .
\ee

The  conditions (\ref{bc1}) specify both $b_n$ and the mass spectrum $m_n$.
There is also a zero mode corresponding to $m_n=0$  in eq.(\ref{eq1}). The (normalized) zero mode solution is easily found to be
\be
f_0=\frac{1}{\pi R}e^{-4a(s-1)|y|}\, . \label{0}
\ee
It should be noted that in order the zero mode to exists, the singular term in eq.(\ref{sin2}) is necessary.
This term   modifies the boundary conditions appropriately and allows the  existence of the zero mode $f_0$.
In particular, if we denote by $S_{\rm bulk}$ the {\it effective} bulk action in an $AdS$ background which describes the dynamics
of the $\phi_{m_1...m_s}$ field, the term which accounts for the singular term in (\ref{sin2}) is
\be  \label{mact}
S=S_{\rm bulk}+ 4a(s\!-\!2) \int d^5x \, \sqrt{-g_{\rm ind}}\,  \delta(y)\, \frac{1}{2} \phi_{m_1...m_s}\phi^{m_1...m_s}\, .
\ee
%This coupling reduces for $s=2$ to the graviton coupling to the brane tension in the standard RS background and
%gives rise to the boundary conditions (\ref{bc0}).
%\be
%f_n'+4 a  (s\!-\!1)\, {f_n}|_{0,\pi R}=0\, . \label{mbc}
%\ee
This extra singular term corresponds to an irrelevant deformation of the boundary $CFT$
 and it has also been proposed in the $AdS$/$CFT$ context in~\cite{Petkou}.

More specifically, the bulk gauge invariant action in ({\ref{mact}), is identical to
 the action  of a bosonic HS field  in an exact $AdS_{5}$ background,
 which turns out to be  \cite{Fronsdal:1978vb}, \cite{BuchBos}
\be \label{hsa1}
S_{\rm bulk}\!\!&=&-\!\!\int d^5x \sqrt{-g}\left(
\frac{1}{2}\nabla_\mu\Phi_{\alpha_1...\alpha_s}\nabla^\mu\Phi^{\alpha_1...\alpha_s}-\frac{1}{2}s
\nabla_\mu{\Phi^\mu}_{\alpha_2...\alpha_s}\nabla_\nu\Phi^{\nu\alpha_2...\alpha_s}\right.\nonumber \\
&&+\frac{1}{2}s(s\!-\!1)
\nabla_\mu{\Phi^\nu}_{\nu\alpha_3...\alpha_s}\nabla^\kappa{\Phi_\kappa}^{\mu\alpha_3...\alpha_s}
-\frac{1}{4}s(s\!-\!1)\nabla_\mu{\Phi^\kappa}_{\kappa\alpha_2...\alpha_s}\nabla^\mu
{\Phi_\lambda}^{\lambda\alpha_2...\alpha_s}\nonumber \\
&&-\frac{1}{8}s(s\!-\!1)(s\!-\!2)
\nabla_\mu{\Phi^{\mu\kappa}}_{\kappa\alpha_4...\alpha_s}\nabla^\nu
{\Phi_{\nu\lambda}}^{\lambda\alpha_4...\alpha_s}\\
&&\!+2a^2\left(s^2\!-s\!-\!4)\right)
\Phi_{\alpha_1...\alpha_s}\Phi^{\alpha_1...\alpha_s} \!-\! a^2 s(s\!-\!1)\!
\left(s^2\!+\! s\!-4\right){\Phi^\mu}_{\mu\alpha_2...\alpha_s}{\Phi_\nu}^{\nu\alpha_2...\alpha_s} \!\!\!
\left. \phantom{\frac{1}{2}}\!\!\!\! \right)\ . \nonumber
\ee
The derivatives are covariant derivatives with respect to the $AdS$ space.
This action is  invariant under the transformation (\ref{df}) in an exact $AdS_{5}$ background, as we shall discuss
in Section $6$. As we have already mentioned,
 one need to commute covariant derivatives in order to prove gauge invariance. These commutations produce Riemann and Ricci tensors, which
in the $AdS$ case are proportional to the metric and can completely be cancelled by the last two terms in (\ref{hsa1}).
 However, in the RS case, there
are additional terms which are not cancelled and emerge from the singular part of the Riemann and Ricci tensors. Denoting these parts
by $\Delta R$ , we have  for example
\be
\Delta {R^5}_{m5n}=-4 a \delta(y) e^{-2 \sigma}\eta_{mn}, ~~~~\Delta R_{mn}=-4 a\delta(y)e^{-2 \sigma}\eta_{mn}, ~~~~
\Delta R_{55}=16 a\delta(y)\ .
\ee
As a result, in the gauge variation of (\ref{hsa1}), there are uncancelled terms proportional to the singular $\Delta R$. Nevertheless, these
contributions can still be cancelled by adding appropriate terms to (\ref{hsa1}). One may prove that indeed, the action
\be
S'=S+S_{\delta}
\ee
is gauge invariant in a RS background for
\be
S_{\delta}=\frac{1}{2}\int d^5x\sqrt{-g}\!\!\!&\!\!\! &\!\!\!\left(
\frac{1}{2}s^2(s\!-\!1)(s\!-\!2){\Phi^{\kappa\mu\alpha_4...\alpha_s}}_\kappa\Delta R^\nu_\mu {\Phi^\lambda}_{\alpha_4...\alpha_s\nu\lambda}\right.
\nonumber \\
&& +s(s\!-\!2)\Phi^{\mu\alpha_2...\alpha_s}\Delta R^\nu_\mu \Phi_{\alpha_2...\alpha_s\nu} \nonumber\\&&
\left. -
 s(s\!-\!2)(s^2\!\!-\!5s\!-\!4)\Phi^{\alpha_1...\alpha_s}\Delta {{R_{\alpha_1}}^{\mu\nu}}_{\alpha_2} \Phi_{\alpha_3...\alpha_s\mu\nu} \right.
 \nonumber \\
 &&\left. -
 \frac{1}{2}s^2 (s\!-\!1)^2(s^2\!\!-\!5s\!-\!4){\Phi^{\kappa\mu\alpha_4...\alpha_s}}_\kappa
 \Delta {{R_{\alpha_1}}^{\mu\nu}}_{\alpha_2} {\Phi^\lambda}_{\alpha_5...\alpha_s\mu\nu\lambda}
\right)\ .
\ee
Obviously, $S_\delta$ is a boundary term as it is proportional to $\delta(y)$.
Then, the transverse traceless part of the HS satisfies eq.(\ref{sin}) and $S_\delta$ reduces to the singular part of (\ref{mact}) as expected.

\section{Half-integer spins in braneworld}

We will now study fermionic fields of half-integer spin $s$ propagating in the bulk of $AdS$ spacetimes.
Such  fields are described by
 totally symmetric tensor-spinors of rank $s\!-\!\frac{1}{2}$, $\Psi_{\alpha_1...\alpha_{s-1/2}}$, and their dynamics
is governed by
the equation
\be\label{general}
 \gamma^\rho\bar\nabla_{\ \rho}\Psi_{\ \alpha_1\ ...\alpha_{s-1/2}}-
 \gamma^\rho\bar\nabla_{(\alpha_1}\Psi_{\ \alpha_2\ ...\alpha_{s-1/2})\rho}
 +\beta
\Psi_{\alpha_1...\alpha_{s-1/2}}=0\ .
\ee
It can straightforward be proven that (\ref{general}) in $AdS_{p+1}$
 is invariant under the gauge  transformation
\be
\delta\Psi_{\alpha_1...\alpha_{s-1/2}}=\bar\nabla_{(\alpha_1}\epsilon_{\alpha_2...\alpha_{s-1/2})}\ ,
\label{generalvar}
\ee
 when the gauge parameter satisfies  the constraint $\gamma^\rho\epsilon_{\alpha_1...\rho...\alpha_{s-3/2}}=0$,
 for the particular value
\be
\beta=(2s-3)\sqrt{\frac{\Lambda}{p}}\, . \label{bb}
\ee
We proceed to solve eq.(\ref{general}) in slices of $AdS_5$ spacetime, where $\beta$ is now given according to
(\ref{bb}) by
\be
\beta=2a (2s-3)\, .
\ee
Note that the $s=1/2,3/2$ cases have been studied in~\cite{Grossman:1999ra},\cite{Pomarol:1999ad},\cite{GPom}.
By using the gauge condition  $\gamma^\rho\Psi_{\rho...}=0$, eq.(\ref{general}) simplifies to
\be
\gamma^\rho D_\rho\Psi_{\alpha_1...\alpha_{s-1/2}}+2as\,\Psi_{\alpha_1...\alpha_{s-1/2}}=0\ .
\ee
We may impose the conditions
\be
\Psi_{\alpha_1...5...\alpha_{s-1/2}}=0=\Psi^\mu{}_{\alpha_1...\mu...\alpha_{s-1/2}}\ ,
\ee
as they are consistent with the field equations (\ref{general}). In the sequel, it is convenient to introduce
 the new fields $\Phi_{a_1...a_{s-1/2}}$, defined by
\be
\Phi_{a_1...a_{s-1/2}}=e^{\sigma(s-5/2)}\Psi_{a_1...a_{s-1/2}}\ ,  ~~~~a_1,a_2,...=0,1,2,3\, .
\ee
Projecting in left/right (L/R) chirality modes, we obtain the following two coupled differential equations
for these fields in $AdS_5$ spacetime
\begin{eqnarray}\label{highspin}
\gamma^c\partial_c\Phi^R_{a_1...a_{s-1/2}}+\partial_5\Phi^L_{a_1...a_{s-1/2}}+2as\Phi^L_{a_1...a_{s-1/2}}&=&0\cr
\gamma^c\partial_c\Phi^L_{a_1...a_{s-1/2}}-\partial_5\Phi^R_{a_1...a_{s-1/2}}+2as\Phi^R_{a_1...a_{s-1/2}}&=&0\ .
\end{eqnarray}
 We can solve the above system by separation of variables
\be
\Phi^{L,R}_{\alpha_1...\alpha_{s-1/2}}=\sum_n f_{L,R}^{(n)}(y)\, \psi^{(n)}_{\alpha_1...\alpha_{s-1/2}}(x^a)\ .
\ee
Recalling that $\gamma^a=e^{\sigma}\hat\gamma^a$ where $\hat\gamma^a$ are the gamma matrices in flat
Minkowski spacetime, we consider the eigenvalue problem
\be
\hat\gamma^a\partial_a\psi^{(n)}_{\alpha_1...\alpha_{s-1/2}}=m_n\psi^{(n)}_{\alpha_1...\alpha_{s-1/2}}\ ,
\ee
which  defines the 4D HS spectrum.
The system of eqs.(\ref{highspin}) reduce then to the first order differential equations
\begin{eqnarray}\label{eqf}
e^{\sigma}m_n f_R+f_L '+2asf_L&=&0\cr
e^{\sigma}m_n f_L-f_R '+2asf_R&=&0\, , \label{s}
\end{eqnarray}
which, in terms of  the new variable $z=\frac{e^{\sigma}m_n}{2a}$,  are written as
\begin{eqnarray}
f_R+\partial_z f_L+\frac{s}{z}f_L&=&0\cr
f_L-\partial_z f_R+\frac{s}{z}f_R&=&0\ .
\end{eqnarray}
The solution of the above equations is
\begin{eqnarray}\label{fspinor}
f_L&=& \frac{z^{1/2}}{N_n}\left(J_{\!-s-\frac{1}{2}}(z)+B_n(m_n) Y_{\!-s-\frac{1}{2}}(z)\right)\ ,\cr
f_R&=&\frac{z^{1/2}}{N_n}\left(J_{\!-s+\frac{1}{2}}(z)+B_n(m_n) Y_{\!-s+\frac{1}{2}}(z)\right)\ ,
\end{eqnarray}
where $J_\nu$ and $Y_\nu$ are the Bessel functions. Moreover, the zero modes of the field, which correspond to $m_n=0$ in (\ref{s}),  are
\begin{eqnarray}\label{massless}
f_L&=&f_L^0 e^{-s\sigma}\ ,\cr
f_R&=&f_R^0 e^{s\sigma}\ .
\end{eqnarray}

The boundary and normalization conditions  fix the mass spectrum of the field and all the parameters of
 the solution (\ref{fspinor}).
The normalization condition for the solutions is chosen such that
\be\label{norm}
\int dy\,  e^{\sigma}f_m f_n=\delta_{mn}\ .
\ee
This is equivalent to the requirement of a  canonically normalized kinetic term for the 4D reduced HS fields
$\psi^{(n)}_{\alpha_1...\alpha_{s-1/2}}$. It can explicitly be written as
\be
N_n^2=
\int_0^{\pi R}\!\!\!\!\!dy\, e^{2\sigma}\!\!\left[ \!J_\nu\!\Big{(}\!\frac{m_n\,e^{2a |y|}}{2a}\!\Big{)}\!\!+\!
\!b_\nu Y_\nu \!\Big{(}\!
\frac{m_n\,e^{2a |y|}}{2a}\!\Big{)}\!\right]^2
\ee
where
$\nu=-s\!-\!\frac{1}{2}$ and $\nu= -s\!+\!\frac{1}{2}$ for the left and right modes, respectively.
Moreover, the  boundary condition $\rm{(i)}$ in eq.(\ref{sbc}) are  written in the present case as
\be
f_R(0^+)=f_L(0^+)\, , ~~f_R(\pi R)=f_L(\pi R)\, ,
\ee
and specifies the parameter $B$ and the masses spectrum. Indeed, we get
\be
&&B(m_n)= \frac{J_{\!-s+\frac{1}{2}}(\frac{m_n}{2a})-J_{\!-s-\frac{1}{2}}(\frac{m_n}{2a})}{Y_{\!-s-\frac{1}{2}}(\frac{m_n}{2a})-
Y_{\!-s+\frac{1}{2}}(\frac{m_n}{2a})}\, , \nonumber \\
&&B(m_n)=B(m_n e^{2a \pi R})\, .
\ee
\begin{figure}
\begin{center}
\includegraphics[width=10cm,height=6cm,angle=0]{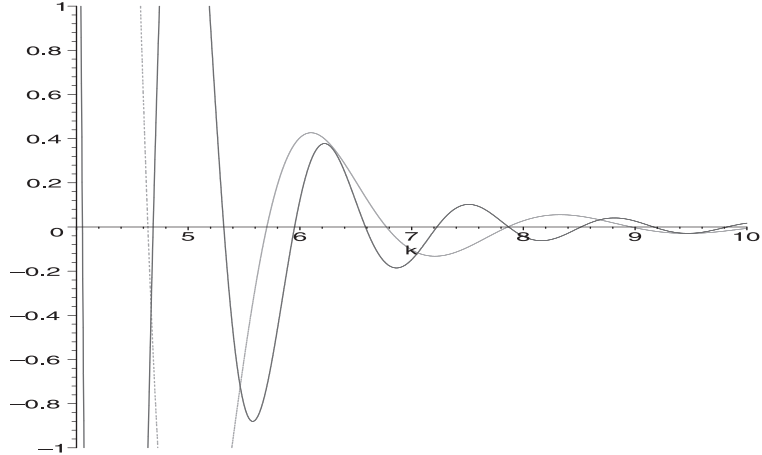}
\end{center}
\caption{Zeros of the function $10^{-2}N_n(f_L-f_R)$ for the simple case of $s=5/2$ with boundary conditions (i).
In the horizontal  axe we used the variable $k=m/a$. The dashed curve is for $R=\ln(5)/2\pi a$. The solid one is for $R=\ln(3)/2\pi a$.
We note that the zeros tend to a continuum for increasing $R$.}\label{fig}
\end{figure}
\begin{figure}
\begin{center}
\includegraphics[width=10cm,height=6cm,angle=0]{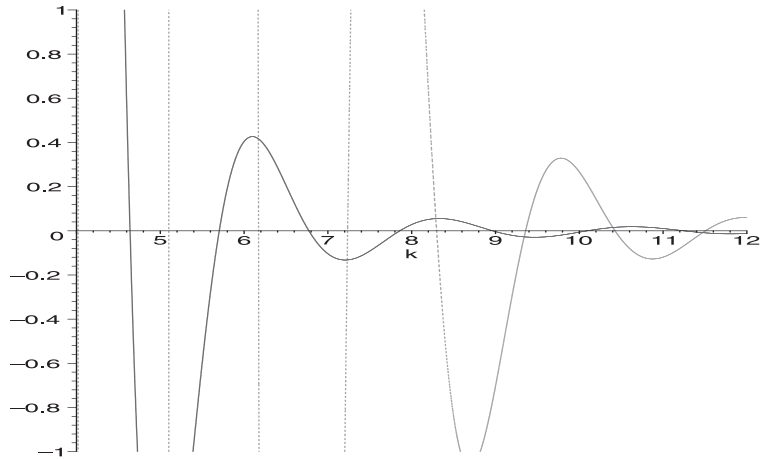}
\end{center}
\caption{Zeros of the function $10^{-2}N_n(f_L-f_R)$ for $R=\ln(3)/2\pi a$ and boundary conditions (i).
Solid line $s=5/2$, dashed $s=7/2$.
We note that increasing the spin means to shift the mass
spectrum further.}\label{fig11}
\end{figure}

There is no  analytical solution for the mass spectrum, but   instead we have
plotted  the function $f_R(\pi R)-f_L(\pi R)$ in fig.(\ref{fig}) and fig.(\ref{fig11}).
The set of zeros correspond to the mass spectrum.
We note that
there exists an infinite tower of massive states for finite $R$. For $R$ increasing the modes become
closer one to each other until the limit of a continuum spectrum
when $R\rightarrow \infty$.
Moreover we note that the massless modes (\ref{massless})
do not satisfy the boundary conditions and thus it is not in the physical spectrum. As a result the spectrum
corresponding to the boundary condition $\rm{(i)}$, which mixes left
and right modes, consists of a tower of massive modes with no massless field.

However, the situation is different for the boundary conditions $\rm{(ii)}$ and $\rm{(iii)}$.
For example, the boundary conditions
$\rm{(iii)}$ give
\be
&&B(m_n)= -\frac{J_{\!-s+\frac{1}{2}}(\frac{m_n}{2a})}{Y_{\!-s+\frac{1}{2}}(\frac{m_n}{2a})}\, , \nonumber \\
&&B(m_n)=B(m_n e^{2a \pi R})\, ,
\ee
from where, the mass spectrum and the constants $B$ are specified. In this case however, there is the zero mode
\be
f_L&=&f_L^0 e^{-s\sigma}\, , ~~~f_R=0\, ,
\ee
localized at the $y=0$ boundary. For the boundary condition $\rm{(ii)}$, left and right modes are
interchanged and the zero mode
is a right handed one localized at the $y=\pi R$ boundary. However, in this case, by moving the boundary
to infinity, the right-handed zero mode
blows up, and thus disappears from the physical spectrum in RS2 due to its non-normalazability.

\section{Consistent gravitational couplings}

We have already noted in the introduction that a free HS theory can consistently be defined in
Minkowski spacetime. We have also stressed that problems appear when minimal interactions are introduced.
For example, if additional gauging, as proposed in \cite{Vass},\cite{Vasss},\cite{Sezgin},
are not allowed, gauge breaking terms proportional to the
Riemann tensor emerge.
These terms are non zero even for on shell graviton  and  therefore tree-level unitarity is lost.
The situation is different for massive HS fields. In this case, without introducing additional gauging,
in order to cancel the gauge breaking terms of the massless theory,
a non-minimal interaction like $\frac{1}{m}\Phi_{\alpha\beta...} {\cal R}^{\alpha\mu\nu\beta} \Phi_{\mu\nu...}$
has been proposed~\cite{Porrati}.
This interaction  cancels hard gauge-breaking terms, i.e., terms that do not vanish
at the massless limit, although gauge invariance is still softly broken due to an explicit mass term.
With the addition of the above non-minimal interaction, the theory is lacking of any
hard breaking terms at linearized level, which could violate tree-level unitarity.
  Hence, tree-level unitarity is restored up to the Planck scale~\cite{Porrati},\cite{Cucchieri:1994tx}.
 The prise paid is the violation of  the equivalence principle due to the introduction of the
 non-minimal interaction terms~\cite{Porrati},\cite{Giannakis}. Although such terms looks odd,
 experience from electromagnetic interactions,
 suggests that the physical requirement is the tree-level unitarity~\cite{Tiktopoulos},\cite{Ferrara}
 rather than minimal coupling. It is  clear of course that the massless
limit for the interactive theory is not defined.

In the case we are considering, we have seen in eqs.(\ref{sin},\ref{general})
 that in order to have gauge invariance in $AdS$, a
non-derivative term is needed. This is something like
the mass term discussed above in the  Minkowskian
case. Similarly, when a gravitational perturbation is switched on,
gauge breaking terms proportional to the Riemann tensor of the
graviton appear again. As in \cite{Porrati}, one can hope that an
equivalent non-minimally coupled interaction may cancel the Riemann
tensor contribution to the variation of the action under the gauge
transformation for the HS field. Contrary to the four
dimensional case in Minkowski background, in our case the
non-derivative term does not break gauge invariance. Therefore the only
cancellation of the hard terms restore the gauge invariance for
the interacting theory, at least at linearized level.
We will show below that it is actually possible to consistently make the interaction
theory gauge invariant at linearized level.
A different perspective has been introduced in \cite{Vasss}
and
it would be interesting to connect this approach with ours. However this is beyond the scope of this paper.

\subsection{Non-minimally coupled lagrangian}

For simplicity, in the following we concentrate on half integer spins, although the discussion can be generalized
along the lines of \cite{Cucchieri:1994tx} to include integer HS fields as well.
For this, let us write the equation of motion (\ref{general}) for a fermionic field
$\Psi_{\alpha_1...\alpha_{s-1/2}}$ of spin $s$  as
\be\label{Q}
Q_{\alpha_1...\alpha_{s-1/2}}=0\, , \label{hse}
\ee
where we have defined
\be
Q_{\alpha_1...\alpha_{s-1/2}}=\gamma^\rho\bar\nabla_{\ \rho}\Psi_{\ \alpha_1\ ...\alpha_{s-1/2}}-
 \gamma^\rho\bar\nabla_{(\alpha_1}\Psi_{\ \alpha_2\ ...\alpha_{s-1/2})\rho}
 +2a (2s-3)
\Psi_{\alpha_1...\alpha_{s-1/2}}\ .
\ee
The action for this field is a generalization of the action in~\cite{Fang:1978wz},\cite{deWit:1979pe} on
Minkowski background and can be written as
\begin{eqnarray}
S=\int d^5x
\sqrt{-g}\Big{[}\!-\!\frac{1}{2}\bar{\psi}_{\alpha_1...\alpha_{s-1/2}}Q^{\alpha_1...\alpha_{s-1/2}}
\!\!&\!+\!&
\!\!\frac{1}{4}(s\!-\!\frac{1}{2})\bar{\psi}_{\mu\rho\alpha_3...\alpha_{s-1/2}}
\gamma^\rho\gamma_\sigma Q^{\sigma\mu\alpha_3...\alpha_{s-1/2}}\cr
\!\!\!&\!\!+\!\!&
\!\!\!\frac{1}{8}(s\!-\!\frac{1}{2})(s\!-\!\frac{3}{2})\bar{\psi}^\mu{}_{\mu\alpha_3...\alpha_{s-1/2}}
Q^\nu{}_\nu{}^{\alpha_3...\alpha_{s-1/2}} \Big{]} \label{as}
\end{eqnarray}
It should be noted that the field equations, which follows from this action, is not eq.(\ref{Q}) but rather
\begin{eqnarray} \label{expanded}
Q_{\alpha_1...\alpha_{s-1/2}}-\frac{1}{2}
\gamma_{(\alpha_1}\gamma^\rho Q_{\alpha_2...\alpha_{s-1/2})\rho}-
\frac{1}{2}g_{(\alpha_1\alpha_2} Q^\mu{}_{\alpha_3...\alpha_{s-1/2})\mu}=0\ .
\end{eqnarray}
However, contraction  with the metric and with a $\gamma$-matrix gives
\be
Q^{\mu...}{}_\mu=0\, , ~~~~~\gamma^\rho Q_{\rho\alpha_2...\alpha_{s-1/2}}=0\, ,
\ee
so that (\ref{expanded}) is actually equivalent to (\ref{Q}).
There is no invariant action which leads directly to (\ref{Q}) without introducing auxiliary
fields~\cite{deWit:1979pe}.

Concerning now the gauge invariance of (\ref{as}), it can be checked by  using the gauge condition
$\gamma^\mu\epsilon_{\mu...}=0$ and the
Majorana-flip identity
\be
\bar{\Psi}_{\alpha_1...\alpha_{s-1/2}}\gamma^{\beta_1...\beta_n}\xi_\lambda=(-)^{n}\bar{\xi}_\lambda
\gamma^{\beta_1...\beta_n}\Psi_{\alpha_1...\alpha_{s-1/2}}\ .
\ee
Then, the variation of (\ref{as}) after an integration by parts turns out to be
\be
\delta S\sim \int d^5x \sqrt{-g}\,\, \bar{\epsilon}_{\alpha_2...\alpha_{s-1/2}}
\!\!\!\!\!\!&\!\!\!\!&\Big{(} D_\mu Q^{\mu\alpha_2...\alpha_{s-1/2}}\!-\!\gamma^{\rho\sigma}D_\rho{Q_\sigma}^{\alpha_2...\alpha_{s-1/2}}
\nonumber \\
&&-D^{(\alpha_2}Q^{\alpha_3...\alpha_{s-1/2})\mu}{}_{\mu}\!-\!2s
\, a \, \gamma_\mu Q^{\mu\alpha_2...\alpha_{s-1/2}}\Big{)}\, . \label{g1}
\ee
By an explicit computation one can show that the integrand in (\ref{g1})
actually vanishes for an $AdS$ background (it is a contracted Bianchi-type
identity).
As a result, the HS theory described by (\ref{as}) is gauge invariant on $AdS$.

Let us now consider possible coupling of the HS fermionic field with gravity in $AdS$ background.
In this case, after performing the gauge variation of (\ref{as}), we will
linearize in the gravitational field $h_{\mu\nu}$ and impose at the end the condition that
both the HS field $\Psi_{\alpha_1...\alpha_{s-1/2}}$ and the graviton are on-shell. Thus,
we will employ the fermionic field equation,
 which in a covariant gauge can be written as
\be
\gamma^\beta{D}_\beta\Psi_{\alpha_1...\alpha_{s-1/2}}+2as\Psi_{\alpha_1...\alpha_{s-1/2}}=0\ ,~~~~\gamma^\mu
\Psi_{\mu\alpha_2...\alpha_{s-1/2}}=0\, ,
\ee
as well as the graviton equation in the $AdS$ background
\be
R_{\mu\nu}(h)=-\Lambda\,  h_{\mu\nu}\, .
\ee
By an explicit evaluation of (\ref{g1}), we then get that
\be\label{delta}
\delta S=2(s-\frac{3}{2})^2\int d^5 x \sqrt{-g}\, \bar{\epsilon}_{\nu\alpha_3...\alpha_{s-1/2}}
\gamma^\mu {W^\nu}_{\beta \mu\lambda}
\Psi^{\lambda \beta\alpha_3...\alpha_{s-1/2}}\ ,
\ee
where ${W^\nu}_{\alpha \mu\lambda}$ is the Weyl tensor. In the $AdS$ case ${W^\nu}_{\alpha \mu\lambda}=0$
and therefore $S$ is
gauge invariant when there is not coupling of the HS to gravity as we noted above.
In fact, the only solutions of Einstein equation in vacuum (including a cosmological constant) with zero Weyl tensor
are maximally symmetric. In this class there are only three possible spacetimes
$(A)dS$ or Minkowski.
Therefore, as soon as a gravitational perturbation is switched on,
the action (\ref{as}) looses its gauge invariance.

In order to maintain gauge invariance of the HS action,
we have to add a term, which  will contain the Weyl tensor  and it will be such that its gauge variation
cancels the  gauge breaking term (\ref{delta}).
 Let us therefore consider the  interaction term
\be\label{first}
\Delta S_1=-\frac{(2s-3)^2}{20a}\int d^5 x \sqrt{-g} \bar\Psi_{\mu\nu\alpha_3...
\alpha_{s-1/2}}{\cal{W}}^{\mu\rho\nu\sigma}\Psi_{\rho\sigma}{}^{\alpha_3...\alpha_{s-1/2}}\ ,
\ee
where $${\cal{W}}^{\mu\rho\nu\sigma}=W^{\mu\rho\nu\sigma}-\frac{1}{2} W_{\alpha\beta}{}^{\rho\mu}
\gamma^{\nu\sigma\alpha\beta}.$$
Another term which can be added and it is  zero on an exact $AdS$ background is
\be\label{second}
\Delta S_2=\frac{(2s-3)^2}{160\, a^3
}\int d^5x \sqrt{-g}\bar\Psi^\lambda{}_{\mu\nu\alpha_4...\alpha_{s-1/2}}\gamma_\lambda
\left[(\alpha\nb+m)\bar\nabla_{\alpha_3}
{\cal{W}}^{ \mu\rho\nu\sigma}\right]\Psi_{\rho\sigma}{}^{\alpha_3...\alpha_{s-1/2}}\ ,
\ee
where $m=2a(s-5/2)$ is the coefficient of the non-kinetic term into the action, when gauge conditions are imposed and
$\alpha$ is a free dimensionless parameter.

Clearly $\Delta S_2$ can only be written for $s\geq 7/2$ and
it does not  contribute to the gravitational multipoles of the spin-$s$ particle as
it can be eliminated by gauge transformations of $\Psi$.

We will now show that the variation of the actions (\ref{first},\ref{second}) exactly cancel the hard term
on shell (which correspond to the first order in the perturbation theory)
(\ref{delta}) up to a local
redefinition of the fields.
Since the identity for a Majorana spinors hold we can just concentrate on the variation of $\bar \Psi$
\be
\delta\bar\Psi_{\mu\nu\alpha_3...\alpha_{s-1/2}}=D_{(\mu}\bar\epsilon_{\nu)A}
-a\bar\epsilon_{A(\mu}\gamma_{\nu)}+(s-5/2)
\left(D_{\alpha_3}\bar\epsilon_{\mu\nu\alpha_4...\alpha_{s-1/2}}-a\bar\epsilon_{\mu\nu\alpha_4...
\alpha_{s-1/2}}\gamma_{\alpha_3}\right)\ ,
\ee
where we introduced the compact notation $(A=\alpha_3...\alpha_{s-1/2})$.
It can easily be proven that the Weyl tensor satisfies
\be
\nabla_\mu W_{\alpha\beta}{}^{\rho\nu}\gamma^{\mu\sigma\alpha\beta}=0\, , ~~~~
\nabla_\mu W^{\mu\rho\nu\sigma}=0\, ,
\ee
thanks to the Bianchi identities. Moreover using the gauge conditions, the
cyclic identity and the fact that $\Psi_{...}$ is a totally symmetric tensor,
 one can prove by direct computation that
\be
W_{\alpha\beta}{}^{\rho[\mu}\gamma^{\nu]\sigma\alpha\beta}D_\mu\Psi_{\rho\sigma A}=0\, .
\ee
Using these identities, after an integration by parts one gets
\begin{eqnarray}
\delta\Delta S_1=-\frac{(2s-3)^2}{10a}\int d^5 x \sqrt{-g}
\!\!\!&\!&\!\!\!\Big{\{}\bar\epsilon_{\nu A} W^{\mu\rho\nu\sigma}\left[D_{[\mu}\Psi_{\rho]\sigma}{}^A
-\gamma^{\epsilon\lambda\mu\rho} D_{\epsilon}\Psi_{\sigma\lambda}{}^A\right]\nonumber \\
&&-(s-5/2)
\{ \bar\epsilon_{\mu\nu\alpha_4...\alpha_{s-1/2}}\bar\nabla_{\alpha_3}
\left({\cal{W}}^{ \mu\nu\rho\sigma}\Psi_{\rho\sigma}{}^{\alpha_3...\alpha_{s-1/2}}\right)\Big{\}}.
\end{eqnarray}
Rearranging the $\gamma$ matrices and considering the equation of motions
$\gamma^\epsilon D_{\epsilon}\Psi_{...}=-2as\Psi_{...}$ one has
\begin{eqnarray}\label{ds1}
\delta\Delta S_1 &=& -2(s-3/2)^2\int d^5 x \sqrt{-g}\, \bar{\epsilon}_{\nu\alpha_3...\alpha_{s-1/2}}
\gamma^\mu {W^\nu}_{\beta \mu\lambda}
\Psi^{\lambda \beta\alpha_3...\alpha_{s-1/2}}+ \cr
&+& \frac{(s-3/2)^2(s-5/2)}{5a}\int d^5 x\sqrt{-g}\bar\epsilon_{\mu\nu\alpha_4...\alpha_{s-1/2}}
\left[\bar\nabla_{\alpha_3}{\cal{W}}^{ \mu\rho\nu\sigma}\Psi_{\rho\sigma}{}^{\alpha_3...\alpha_{s-1/2}}\right]\ .
\end{eqnarray}
The first line of (\ref{ds1}) cancel the hard term (\ref{delta}). The total variation turns out then to be
\begin{eqnarray}
\delta S+\delta\Delta S_1+\delta\Delta S_2&=&-\frac{(2s-3)^3}{80a^3}
\int d^5 x\sqrt{-g}\bar\epsilon_{\mu\nu\alpha_4...\alpha_{s-1/2}}
\times\cr
&\times & \left[\alpha(\nb^2-m^2)+{\cal{M}}(\nb+m)\right]
\{\bar\nabla_{\alpha_3}{\cal{W}}^{\mu\rho\nu\sigma}\Psi_{\rho\sigma}{}^{\alpha_3...\alpha_{s-1/2}}\}\ ,
\end{eqnarray}
where ${\cal{M}}=2a\left[s(2\alpha+1)-(4\alpha+5/2)\right]$.
Under a local redefinition of $\Psi$ the second line vanish on shell as it is proportional
to the equation of motion for $\Psi$.
As a result, the action $\delta S\!+\!\delta\Delta S_1\!+\!\delta\Delta S_2$ is gauge invariant
for on-shell interacting HS fields
and gravitons.

\subsection{Coupling to brane matter}

The HS fields may also couple to matter
living on the boundary branes. In order to find these couplings, we note that
the interaction term (\ref{first}) induces a boundary action
when the variation of the metric vanishes at the
boundary but not its orthogonal derivative. More explicitly, in
gaussian normal coordinates, the boundary action appears whenever
\be
\delta g_{ab}=0 ~~~~\mbox{and} ~~~~\delta\partial_y g_{ab}=2\delta K_{ab}\neq 0. \label{cond}
\ee
This is the case, for example, in the RS scenario.
Then,  when (\ref{cond}) is valid, by employing the gauge conditions
$\Psi_{5...}=0$ and the Majorana-flip identity, $\Delta S_1$ in (\ref{first}) reduces to
\be\label{var}
\!\!\Delta S_1=\!-\!\frac{(s\!-\!3/2)^2}{5a}\int d^5 x \sqrt{-g} \bar\Psi_{a_1 a_2 a_3...
a_{s\!-\!1/2}}\left[W^{a_1b_1a_2b_2}\!-\!\frac{1}{2}W_{cd}{}^{b_1 a_1}\gamma^{a_2b_2cd}\right]
\Psi_{b_1 b_2}{}^{a_3...a_{s\!-\!1/2}}.
\ee
We recall that the Weyl tensor is expressed in terms of the curvature tensors as
\be
W_{abcd}=R_{abcd}-\frac{1}{3}\left(g_{a[c}R_{d]b}-g_{b[c}R_{d]a}\right)+\frac{1}{12}R g_{a[c}g_{d]b}\ .
\ee
In Gaussian-normal coordinates, only the Ricci tensor and scalar contain a
term proportional to the derivative of the extrinsic curvature \cite{gr-qc/0004021}.
In particular
\be
R_{ab}=\partial_y K_{ab}+...\ .
\ee
By direct computation one can now show that the variation of the action (\ref{var}) with respect to the metric
is defined if and only if the following boundary term is added
\be
S_b=-\frac{2(s-3/2)^2}{5a}\int_b d^4 x \sqrt{-g_{ind}} \bar\Psi^a{}_{a_2...
a_{s-1/2}}\left[K_{ab}-\frac{1}{4}g_{ab}K\right]\Psi^{b a_2...a_{s-1/2}}\ .
\ee
Note that, even if $\Psi_R(0^+)=-\Psi_R(0^-)$ we have that $S_b(0^+)=S_b(0^-)$ since $K(0^+)=-K(0^-)$.

In the case when $K_{ab}\Big|_{0,\pi R}\propto g_{ab}\Big|_{0,\pi R}$, the boundary action vanishes.
This happens for example for the RS scenario. Here in fact only a boundary
mass for the graviton is added. A second important thing to note is that only the massive modes
which satisfy the boundary conditions (i) make the boundary action non vanishing.
In fact, we have proved
that the massless mode, if exists, is chiral and the boundary action mix right with left-handed modes.

If we now allow matter on the brane,
the Israel-Darmois junction conditions relate the extrinsic curvature to the matter content on the
brane as \cite{shiromitzu}
\be
K_{ab}(0^+)=-\frac{1}{2}k^2_5 T_{ab}(0)+...\ ,
\ee
where the dots indicate terms proportional to the induced metric, $k^{-2/3}_5$ is
 the five dimensional Planck mass and $T_{ab}$
the energy momentum tensor of the boundary matter.
Then an effective coupling between the HS-fields, treated as test fields, and matter on the brane appears, which may explicitly be written as
\be
S_b=k_5^2\frac{(s-3/2)^2}{5a}\int_b d^4 x \sqrt{-g_{ind}} \bar\Psi^a{}_{a_2...
a_{s-1/2}}\left[T_{ab}-\frac{1}{4}g_{ab}T\right]\Psi^{ba_2...a_{s-1/2}}\ .
\ee
Note that the boundary description we have presented here will break down when,
using the RS fine tuning \cite{shiromitzu},
$T_{ab}\sim a k_5^{-2}>10 {\rm TeV}^4$, where this limit is compatible with table-top tests of Newton's law (see \cite{germani} and references
therein).

\section{Holography}

In this section we will discuss HS fields in the $AdS$/$CFT$ setup and their holographic interpretation. In particular,
to make a contact with previous literature, we will explore bosonic
HS fields in the standard $AdS_5$ case and fermionic HS fields in RS background.

\subsection{Holography: bosons in $AdS$}

Here, we will  explicitly  calculate the  two-point function of higher-spin operators in the boundary $CFT$ .
 The conformal dimension of   HS operators, in the light-cone formalism, have been calculated in~\cite{metsaev}.
Here, using~\cite{Witten},  the full two-point function of HS bosonic operators will be found
including the tensorial structure. It should be noted that the $s=0,1$ cases have been evaluated initially
in~\cite{Witten}, whereas the $s=2$ one in~\cite{s2}. More references can be found in~\cite{malda}.
In general, the action for a bosonic HS fields in $AdS_{p+1}$ is \cite{BuchBos}
\be \label{hsa}
S\!\!&=&-\!\!\int d^{p+1}x \sqrt{-g}\left(
\frac{1}{2}\nabla_\mu\Phi_{\alpha_1...\alpha_s}\nabla^\mu\Phi^{\alpha_1...\alpha_s}-\frac{1}{2}s
\nabla_\mu{\Phi^\mu}_{\alpha_2...\alpha_s}\nabla_\nu\Phi^{\nu\alpha_2...\alpha_s}\right.\nonumber \\
&&+\frac{1}{2}s(s\!-\!1)
\nabla_\mu{\Phi^\nu}_{\nu\alpha_3...\alpha_s}\nabla^\kappa{\Phi_\kappa}^{\mu\alpha_3...\alpha_s}
-\frac{1}{4}s(s\!-\!1)\nabla_\mu{\Phi^\kappa}_{\kappa\alpha_2...\alpha_s}\nabla^\mu
{\Phi_\lambda}^{\lambda\alpha_2...\alpha_s}\nonumber \\
&&-\frac{1}{8}s(s\!-\!1)(s\!-\!2)
\nabla_\mu{\Phi^{\mu\kappa}}_{\kappa\alpha_4...\alpha_s}\nabla^\nu
{\Phi_{\nu\lambda}}^{\lambda\alpha_4...\alpha_s}\\
&&\!+\!\frac{\Lambda}{2 p}\left(s^2\!+\!(p\!-\!5)s\!-\!2(p\!-\!2)\right)
\Phi_{\alpha_1...\alpha_s}\Phi^{\alpha_1...\alpha_s} \!-\!\frac{\Lambda}{4 p}s(s\!-\!1)\!
\left(s^2\!+\!(p\!-\!3) s\!-p\right){\Phi^\mu}_{\mu\alpha_2...\alpha_s}{\Phi_\nu}^{\nu\alpha_2...\alpha_s} \!\!\!
\left. \phantom{\frac{1}{2}}\!\!\!\! \right)\ . \nonumber
\ee
where the derivatives are covariant derivatives in the $AdS$ background.
The field equations which follows from eq.(\ref{hsa}) are
\be
&&\nabla^2\Phi_{\alpha_1...\alpha_s}-\nabla_{(\alpha_1}\nabla^\mu\Phi_{\alpha_2...\alpha_s)\mu}
+\frac{1}{2} \nabla_{(\alpha_1}\nabla_{\alpha_2}{\Phi^\mu}_{\alpha_3...\alpha_s)\mu}
+ g_{(\alpha_1\alpha_2}\nabla_\mu\nabla_\nu{\Phi^{\mu\nu}}_{\alpha_3...\alpha_s)}-
\nabla_{(\alpha_1}\nabla_\mu\Phi^\mu_{\alpha_2...\alpha_s)}-\nonumber \\
&&
-g_{(\alpha_1\alpha_2}\nabla^2{\Phi^{\mu}}_{\alpha_3...\alpha_s)\mu}-
\frac{1}{2}g_{(\alpha_1\alpha_2}\nabla_{\alpha_3}\nabla^\mu{\Phi^{\nu}}_{\alpha_4...\alpha_s)\mu\nu}
-\frac{\Lambda}{p}\left(s^2+(p-5)s-2(p-2)\right)\Phi_{\alpha_1...\alpha_s}\nonumber \\
&&+\frac{\Lambda}{p}\left(s^2+(p-3) s-p\right)
g_{(\alpha_1\alpha_2}{\Phi^{\mu}}_{\alpha_3...\alpha_s)\mu}=0
\ee
The transverse traceless part of the HS fields
$
\nabla^\mu\Phi_{\mu...}={\Phi^{\mu}}_{\mu...}=0
$
 satisfy the free wave equation
\be
\nabla^2\Phi_{\alpha_1...\alpha_s}-\frac{\Lambda}{p}\left(s^2+(p-5)s-2(p-2)\right)
\Phi_{\alpha_1...\alpha_s}=0 \, . \label{eqq}
\ee
We remind again that the parenthesis in the indices denote symmetrization
without combinatorial factors (i.e, $A_{(\mu}B_{\nu)}= A_{\mu}B_{\nu}+
A_{\nu}B_{\mu}$).
For later use, we note that when the equations of motions are obeyed,
the action (\ref{hsa}) turns out to be the total divergence
\be
S=-\int d^{p+1}x \sqrt{-g} \, \nabla_\mu V^\mu  \, , \label{aaa}
\ee
where
\be
V^\mu&=& \frac{1}{2}\Phi^{\alpha_1...\alpha_s}\nabla^\mu\Phi_{\alpha_1...\alpha_s}-\frac{1}{2} s \Phi^{\mu\alpha_2...\alpha_s}\nabla^\nu \Phi_{\nu\alpha_2...\alpha_s}-\frac{1}{4}
s(s-1)\Phi^{\mu\nu\alpha_3...\alpha_s}\nabla_\nu {\Phi^\lambda}_{\lambda\alpha_2...\alpha_s}\nonumber \\
&&
-\frac{1}{4}
s(s-1){\Phi^\nu}^{\nu\alpha_3...\alpha_s}\nabla^\mu {\Phi^\lambda}_{\lambda\alpha_2...\alpha_s}
-\frac{1}{8}
s(s-1)(s-2){\Phi^{\mu\kappa}}_{\kappa\alpha_4...\alpha_s}\nabla^\nu {\Phi_{\nu\lambda}}^{\lambda\alpha_4...\alpha_s}\nonumber \\&&
+\frac{1}{4}
s(s-1){\Phi^\nu}_{\nu\alpha_3...\alpha_s}\nabla^\lambda {\Phi_\lambda}^{\mu\alpha_3...\alpha_s} \, . \label{vvv}
\ee

In the following we will restrict ourselves in the  $AdS_5$ case so that $p=4$, although the discussion may be kept more general.
We will employ the  conformally flat Poincar\'e coordinates for $AdS_5$ so that the metric is written as
\be\label{metriconf}
ds^2=\frac{1}{4a^2x_0^2}(dx^a dx_a+dx_0^2)\ , \label{poin}
\ee
moreover we will make use of the Euclidean signature.

There are two boundaries  $x_0=0$ and $x_0=\infty$. The $x_0=0$ boundary  is the 4D Minkowski spacetime whereas the
$x_0=\infty$ one is actually  a point as all the
four dimensional points are shrank to zero. Thus, the boundary of
$AdS_5$ is the 4D compactified Minkowski spacetime (Minkowski plus the point at
infinity).
 To extract different four dimensional points from the boundary at infinity one can make an $SO(2,4)$
transformation that map the point $x_0=\infty$ to $(x_0=0,x^a)$ and leave invariant the boundary at $x_0=0$.
This transformation is an isometry for $AdS_5$ and correspond just
to a conformal transformation on the Minkowskian boundary $x_0=0$. Our aim is to find the
function $\Phi$ at the boundary in terms only  of
the boundary data $\phi_{a_1...a_s}$ at $x_0=0$. We therefore look for a kind of propagator
(Green function) for the field $\phi$ at the boundary.
Since the point at infinity is mapped to the point at zero, it is much simpler to find a
divergent solution of $\Phi$ at infinity and then map the
point to zero. However, one has to be careful in taking the limit $x_0=0$ as some divergences may appear.
We therefore consider a boundary
on a point $x_0=\epsilon$ and then we take the limit $\epsilon\rightarrow 0$. Such a
limit is finite and the limit process may be
interpreted as a renormalization process.

We will consider the holographic interpretation for the massless higher spin field.
The massless mode is going to be mapped to a
boundary conformal invariant operator. Concerning the massive ones,
a massive KK mode in $d=p+1$ dimensions behaves at the boundary $x_0=\infty$ equivalent to $y\rightarrow \infty$ as
\be
\Phi_m\sim x_0^{(p-1)-2s}\, .
\ee
 Therefore all the massive modes for a spin field $s>(p-1)/2$ do not contribute at the boundary.
In five dimensions the only massive mode that could contribute are  $s=1$ gauge and $s=0$ scalar fields.
However, the effect of massive KK  state is to introduce
logarithmic divergences, which  can be
reabsorbed by  renormalization \cite{malda}.
Therefore, the important
modes are only the massless ones.

The next step is to solve the HS field equations and plug back the solution into the action (\ref{hsa}) in order to
calculate the two-point function of HS operators of the boundary $CFT$. For this, we will assume appropriate boundary
conditions, which are written as
\be
\Phi_{a_1...a_s}(x_0=0,x^a)=\phi_{a_1...a_s}(x^a)\,, ~~~~~\Phi_{0 \alpha_2...\alpha_s}(x_0=0,x^a)=0
\ee
where Greek and Latin indices run over 5 and 4 dimensions, respectively ($\alpha_1..=0,...,4,~~ a_1...=1,...,4$).
Moreover, the solutions we are after,  approach a $\delta$-function at the boundary. As in the cases already
discussed~\cite{Witten},\cite{s2}, this can be achieved as follows.  One finds first solutions
which behave like $\delta$-function at
$x_0=\infty$ in the sense that the boundary condition
\be
\Phi_{a_1...a_s}(x_0\to \infty,x^a)\to \infty\, , ~~~~~\Phi_{0 \alpha_2...\alpha_s}(x_0\to \infty,x^a)
=0\, , \label{ash}
\ee
are satisfied and then use the inversion transformation (\ref{inv}) given below,  for the solution at $x_0=0$.
Using this method, one finds that $\Phi_{0\alpha_2...\alpha_s}$ do not couple to
$\Phi_{a\alpha_2...\alpha_s}$, and by  (\ref{ash}),   $\Phi_{0\alpha_2...\alpha_s}$ can be consistently put to
zero ($\Phi_{0\alpha_2...\alpha_s}=0$).
Moreover, we recall that a massless state solution can be written as
\be
\Phi_{a_1...a_s}(x_0,x^a)=f(x_0)\phi_{a_1...a_s}(x^a)\ ,
\ee
where $\phi_{a_1...a_s}(x^a)$  is the boundary value of the field $\Phi$ and it is transverse and traceless
$\nabla^a\phi_{aa_2...a_s}={\phi^a}_{aa_3...a_s}=0$.
Inserting this ansatz in (\ref{eqq})  one find two possible solutions
\be
f_1(x_0)=(2ax_0)^{2(1-s)}\ ,\ f_2(x_0)=(2ax_0)^2\ .
\ee
For $s> 2$, the first solution is the normalizable one and it has been discussed previously. However since we are looking for a divergent solution
on the $x_0=\infty$ boundary we will use the non normalizable one. So the solution is then
\be
\Phi_{a_1...a_s}=N 4a^2x_0^2\phi_{a_1...a_s}(x^a)\ ,
\ee
where $N$ is a normalization factor.

We now map the point at infinity (which make the field divergent)
to a point in zero with the $SO(2,4)$ transformation
\be
x^{\mu}\rightarrow\frac{x^\mu}{x_0^2+\mid x\mid^2}\ , \label{inv}
\ee
where $\mid x\mid$ is the distance of a four dimensional point from the origin. We also introduce the function $\Omega=x_0^2+| x|^2$.
With this transformation we obtain
\be
\Phi_{\alpha_1...\alpha_s}=N 4a^2\frac{x_0^2}{\Omega^{s+2}}{\cal I}_{\alpha_1 b_1}...{\cal I}_{\alpha_s b_s}\phi^{b_1...b_s}\ ,
\ee
where
\be
{\cal I}_{\mu \nu}=\eta_{\mu\nu}-\frac{2 x_\mu x_\nu}{x_0^2+\mid x\mid^2}\ ,
\ee
and all indices are rise and lowered by the Euclidean metric $\eta_{\mu\nu}$.

Even if the function $x_0^2/\Omega^{s+2}$ is divergent in the point $x_0=0=\mid x\mid$ is not yet actually a Dirac delta function so it cannot
represent a Green function for $\phi$. It is very simple to see that instead a Dirac delta function can be represented as
\be\label{dirac}
\delta^{(4)}(x^a)=\lim_{x_0\rightarrow 0}\frac{\pi^2}{s(s+1)}\int d^4 x \frac{x_0^{2s}}{\Omega^{s+2}}\ .
\ee
We can obtain a Green function then by raising $s-1$ indices on $\Phi$ and setting $N=s(s+1)/\pi^2 (2a)^{2s}$. Then a generic
field $\Phi(0,x^{a})$ can be obtained in the limit $x_0\rightarrow 0$ by the superposition
\be\label{super}
\Phi_{\alpha_1}{}^{\alpha_2...\alpha_s}=\frac{s(s+1)}{\pi^2}\int d^4 x'
\frac{x_0^{2s}}{\Omega(x_0,\mid x -x'\mid)^{s+2}}{\cal I}_{\alpha_1b_1}{\cal I}^{\alpha_2}{}_{b_2}
...{\cal I}^{\alpha_s}{}_{b_s}
\phi^{b_1...b_s}(x')\ ,
\ee
where it has to be considered ${\cal I}_{ab}={\cal I}_{ab}(x_0,x-x')$.

From the superposition (\ref{super}) we can always lower and rise index in such a way that
\be
\Phi^{\alpha_1}{}_{\alpha_2...\alpha_s}\!=\!\frac{s(s\!+\!1)}{4\, a^2\pi^2}\int d^4 x'
\frac{x_0^{4}}{\Omega(x_0,\mid x -x'\mid)^{s+2}}{\cal I}^{\alpha_1b_1}{\cal I}_{\alpha_2}{}_{b_2}...
{\cal I}_{\alpha_s}{}_{b_s}
\phi^{b_1...b_s}(x')\ .
\ee
We may also easily calculate the derivative, to be used below, which is given by
\be
\partial_0\Phi^{ \alpha_1}{}_{\alpha_2...\alpha_s}\!=\!\frac{s(s\!+\!1)}{a^2 \pi^2}\int d^4 x'
\frac{x_0^{3}}{\Omega(x_0,\mid x -x'\mid)^{s+2}}{\cal I}^{\alpha_1b_1}{\cal I}_{\alpha_2}{}_{b_2}...{\cal I}_{\alpha_s}{}_{b_s}
\phi^{b_1...b_s}(x') + O(x_0^4).
\ee

The next step is to evaluate the action (\ref{hsa}) for the field we found above. Taking into account that the action (\ref{hsa}) is
written as a total derivative (\ref{aaa}), evaluation of (\ref{vvv}) gives
\be
S_B=-\int d^4 x \sqrt{-g_{\rm ind}}\, \Phi^{a_1...a_s}\partial^0 \Phi_{a_1...a_s}
\ee
The boundary action at $x_0=\epsilon$ is calculated then to be
\begin{eqnarray}
S_B&=&-\frac{s^2(s+1)^2}{a^2 \pi^4}\int d^4 x \epsilon^{-3}\int d^4 x^{'} \int d^4 x^{''}
\frac{\epsilon^{3}}{\Omega(\epsilon,\mid x -x^{'}\mid)^{s+2}}
{\cal I}_{a_1b_1}...{\cal I}_{a_s b_s}
\phi^{b_1...b_s}(x^{'})\times \cr
&\times &\frac{\epsilon^{2s}}{\Omega(\epsilon,\mid x -x^{''}\mid)^{s+2}}
{\cal I}_{c_1d_1}...{\cal I}_{c_s d_s}
\phi^{d_1...d_s}(x^{''})\eta^{a_1c_1}...\eta^{a_sc_s}+O(\epsilon^{2s+1})
\end{eqnarray}
In the limit $\epsilon\rightarrow 0$, recalling the definition of the Dirac delta function (\ref{dirac}) and using the fact that
$\lim_{x\rightarrow 0}{\cal I}_{\mu\nu}=\eta_{\mu\nu}$ keeping $\epsilon\neq 0$, we obtain
\begin{eqnarray}\label{re}
S_B=-\frac{s(s+1)}{a^2\pi^2}\int\int d^4 x d^4 x^{'}
\frac{\phi^{a_1...a_s}(x){\cal I}_{a_1b_1}...{\cal I}_{a_s b_s}
\phi^{b_1...b_s}(x^{'})}{\mid x-x^{'}\mid^{2s+4}}\ .
\end{eqnarray}
As $\phi$ is symmetric and traceless and ${\cal I}_{a_1(b_1}{\cal I}_{b_2)a_2}$ is completely symmetric,
the action (\ref{re}) may be rewritten as
\begin{eqnarray}\label{kernel}
S_B=-\frac{s(s+1)}{a^2\pi^2}\int\int d^4 x d^4 x^{'}
\frac{\hat\phi^{a_1...a_s}(x){\cal E}_{a_1...a_s}{}^{c_1...c_s}{\cal I}_{c_1b_1}...{\cal I}_{c_s b_s}
\hat\phi^{b_1...b_s}(x^{'})}{\mid x-x^{'}\mid^{2s+4}}\ ,
\end{eqnarray}
where $\hat \phi$ is any initial condition at the boundary $x_0=0$ and ${\cal E}_{a_1...a_s}{}^{c_1...c_s}$ is the projector onto totally
symmetric traceless s-index tensor defined as \cite{Osborn}
\be
\!\!{\cal E}_{a_1...a_s}{}^{c_1...c_s}\!=\! \frac{\delta_{(a_1}{}^{(c_1}...\delta_{a_s)}{}^{c_s)}}{(s!)^2}
\!+\!\frac{1}{s!}\!\sum^{[s/2]}_{r=1}\!\lambda_r g_{(a_1 a_2}...g_{a_{2r-1}a_{2r}}
g^{(c_1c_2}\!\!...g^{c_{2r-1}c_{2r}}\!\delta_{a_{2r+1}}{}^{c_{2r+1}}\!\!...\delta_{a_s)}{}^{c_s)}\ ,
\ee
where $[s/2]$ is the integer part of $s/2$ and
\be
\lambda_r=(-1)^r\frac{1}{2^r r!(s-2r)!\prod_{k=1}^r(4+2s-2-2k)}\ .
\ee
Let us now consider a lagrangian in four dimension
\be
{\cal L}={\cal L}_{CFT}+\hat\phi_{a_1...a_s}J^{a_1...a_s}+...
\ee
where $\hat\phi$ is an external frozen field and $J_{a_1...a_s}$ is a conserved and traceless current of dimension $(s+2)$ of the $CFT$.
Then one has \cite{Anselmi}
\be\label{<>}
\langle J_{a_1...a_s}(x) J_{b_1...b_s}(x^{'})\rangle\sim\frac{{\cal E}_{a_1...a_s}{}^{c_1...c_s}{\cal I}_{c_1b_1}...
{\cal I}_{c_s b_s}}
{\mid x-x^{'}\mid^{2s+4}}\ .
\ee
The equation (\ref{<>}) is equivalent to the kernel of (\ref{kernel}) in accordance with the $AdS$/$CFT$ correspondence.

The scenario described above, change completely when one looks at the RS model.
First of all the ``visible'' boundary is not anymore at spatial
infinity and Neumann boundary conditions must be imposed. Thanks to the boundary
conditions, the fields can also be dynamical in the holographic picture
and the tower of massive modes do not decay.
In this way the KK modes contribute to the holographic
theory switching on non conformally invariant operators on the
holographic theory at the boundary. However, removing the IR brane, the conformal invariance is restored and the
conformal dimension of the dual operators equal their bare dimensions.

In the following we will discuss this case for fermionic fields but the bosonic case can be directly generalized from it.

\subsection{Holography: fermions in a Box}

We will discuss here the holographic picture of higher
spins fermions in the RS model on the UV brane. The tensorial structure for fermionic operators is similar
to  the bosonic ones, and thus, we will only consider their scaling properties. Note that the case $s=1/2$ has
been already discussed by several authors \cite{CoPo}.

We are interested in computing the two-point function of higher-spin operators in the RS case.
We will again use Poincar\'e coordinates with metric as in (\ref{poin}).
Following closely  analogous computations for the lowest spin cases in standard $AdS$/$CFT$ and in RS, we will
put the UV brane, the UV regulator, at $x_0= 1/2a$  and the $\rm{TeV}$ brane at $x_0=1/\mu$.
We are interested in the large $N$ limit of the corresponding holographic theory which is equivalent
of requiring a large cosmological constant or, in particular, a small $x_0$.

The CFT  is living in the UV
boundary and as fixed source fields will be taken the left-handed
HS fermionic field defined by the conditions
(suppressing  tensor indices for convenience)
\be \Psi_L(x_0=\frac{1}{2a},x^a)=\Psi^0_L(x^a) \, ,~~~~~
\mbox{with}~~~~~~~  \delta\Psi_L\Big{|}_{1/2a}=0,~~~~
\Psi_L\Big{|}_{1/\mu}=0 \, ,
\ee
whereas, the right-handed component $\Psi_R$ will
be free. The fermionic HS action is given by eq.(\ref{as}) and its
variation does not vanish as the right-handed HS fields are free
on the UV boundary. Thus, we are forced to add a boundary term. In
the $\gamma^\mu \Psi_{\mu...}=0$ gauge, this is
\be
S_{\rm boundary}&=&\frac{1}{g_5^2}\int_{UV} d^4 x \sqrt{-g_{ind}}
\left(\frac{1}{2}\bar{\Psi}_{\alpha_1...\alpha_{s-1/2}}{\Psi}^{\alpha_1...\alpha_{s-1/2}}
\right) , \label{bt}
\ee
where $g_5^2$ is the bulk coupling constant of the fermionic gauge field $\Psi_{\alpha_1...\alpha_{s-1/2}}$.

We recall
that in the gauge  $\gamma^\mu{\Psi}_{\mu...}=0$, the HS fermionic
field equations turn out to be
\be \D
\Psi_{\alpha_1...\alpha_{s-1/2}}+2as \Psi_{\alpha_1...\alpha_{s-1/2}}=0\
, \ee
as $\Psi_{5...}$ decouples and can consistently be taken to
vanish. These equations, after  projecting with $1\pm\gamma^5$ are
written as
\begin{eqnarray}\label{two}
&&\partial_0
\Psi_L+(2s-\frac{5}{2})\frac{1}{x_0}\Psi_L+\hat\gamma^a\partial_a\Psi_R=0
\cr &&\partial_0
\Psi_R-\frac{5}{2}\frac{1}{x_0}\Psi_R-\hat\gamma^a\partial_a\Psi_L=0\ ,
\end{eqnarray}
where, for convenience, all tensor indices are suppressed.
Let us consider a solution in the four dimensional momentum space
of the type
\be
\Psi_{L,R}(p,x_0)=\frac{f_{L,R}(p,x_0)}{f_{L,R}(p,1/2a)}\Psi^0_{L,R}(p)\
, \ee
where $\Psi^0_{L,R}(p)$ is the wave function at the UV
boundary. With this separation (\ref{two}) satisfy the equations
\be && \partial_0 f_L+(2s-\frac{5}{2})\frac{1}{x_0}f_{L}-p
f_{R}=0\cr &&
\partial_0 f_R-\frac{5}{2}\frac{1}{x_0}f_{R}+p
f_{L}=0\cr && i \, \p \Psi_{R,L}^0=-p\
\frac{f_{R,L}(p,1/2a)}{f_{L,R}(p,1/2a)}\Psi_{L,R}^0\ . \ee
It is not difficult to verify that the solution for $f_{L,R}$ using the boundary condition
$\Psi_R(1/\mu)=0$ is
\begin{eqnarray}
f_{L}(p,x_0)&=&x_0^{3-s}\left[J_{s+1/2}(p\,x_0)Y_{s-1/2}(p/\mu)-J_{s-1/2}(p/\mu)Y_{s+1/2}(p\,x_0)\right]\cr
f_{R}(p,x_0)&=&x_0^{3-s}\left[J_{s-1/2}(p
\,x_0)Y_{s-1/2}(p/\mu)-J_{s-1/2}(p/\mu)Y_{s-1/2}(p\,x_0)\right]\ .
\end{eqnarray}

It is clear that due to the field equations (\ref{hse}), the
bulk HS action (\ref{as}) vanish on shell and thus the only
contributions results from the boundary term (\ref{bt}). As a result, the
boundary action turns out to be
\be
S_{\rm boundary}=\frac{1}{ g_5^2}\int d^4
p\bar{\Psi}^{0}_L(p)\Sigma(p)\Psi^0_L(-p)\ , \label{acc}
 \ee
 where, we have defined
 \be
~~~~~~\Sigma(p)=-\frac{1}{2}\frac{p}{i\, \p}
\frac{f_R(p,1/2a)}{f_L(p,1/2a)}\ . \label{fac}
\ee
Then, according to the standard nomenclature, we have that $S_{\rm boundary}$ is the generating functional of
connected Green functions in the boundary theory. As a result, we will have
\be
\langle {\cal O}_{a_1...a_{s-1}}(p){\cal \bar{O}}^{c_1...c_{s-1}}(-p)\rangle ={\cal E}_{a_1...a_s}{}^{c_1...c_s}
 g_5^{-2} \Sigma(p)\ \ ,
\ee
where the tensorial structure is encoded in ${\cal E}_{a_1...a_s}{}^{c_1...c_s}$. In the $a\to \infty$ limit,
we get
\be
\langle {\cal O}(p)\, \bar{\cal{O}}(-p)\rangle =
 \frac{-i\,  g_5^{-2}}{2^{2s}\Gamma(s\!+\!\frac{1}{2})} \,\frac{\p}{2a} \, \left(\frac{p}{2a}\right)^{2s-1}
 \Big{(} \ln(p/2a)-
 \pi \frac{Y_{s-1/2}(p/\mu)}{J_{s-1/2}(p/\mu)}\Big{)} \ \ .
\ee
In the above, we have kept only  the first non-analytic term, we have ignored analytic terms and
the tensorial structure has been suppressed.
For Euclidean momenta $p\rightarrow i p$ and $p>>\mu$, we get in particular
\be\label{hsO}
\langle {\cal O}(p) \bar{\cal{O}}(-p)\rangle =
\frac{(-)^{s\!-\!\frac{1}{2}}\,  g_5^{-2} }{2^{2s}\Gamma(s\!+\!\frac{1}{2})}\,
\,\p \, p^{2s-1}  (2a)^{-2s+2}\ln(p/2a) \ \ ,
\ee
which is what is expected for the two-point function of operators of dimension ${\rm dim}[{\cal O}]=s+2$.

Note that in the bosonic sector of the dual $CFT$ to a non-critical string theory with a UV cut-off,
a similar structure to (\ref{hsO})
arises whenever one consider a scalar field with angular momentum $\ell$ \cite{gubser}. In that case the dimension
of the operator ${\cal O}$ of the dual theory is related to the partial wave of momentum $\ell$ of the bulk scalar field,
where again the non-perturbative dimensions of the $\cal O$s equal their bare dimensions.

\section{Conclusions}

It is known, that propagation of free HS fields can consistently be
defined on $AdS$ spacetimes. Here,
we have discussed the dynamics of
such fields not in the whole of $AdS$ but rather in a part of it, and in particular,
in a RS background. The aim was to find the spectrum of the resulting 4D HS theory. To reach this purpose,
we first specified the boundary conditions that should be satisfied by the HS fields living in the bulk of the
$AdS$ spacetime, and then we solved the HS
field equations in the bulk of $AdS$ for all spins,
integer and half-integer. The resulting 4D spectrum consists of an
infinite tower of  massive states. In addition,
there exists a massless mode for spin $s=1$. Massless mode also exist for
bosons with $s>1$ if a boundary term
is introduced. This is a boundary mass for the HS
fields. Hence, with the addition of such term, the
4D spectrum consist of an infinite tower of massive as well as massless modes for all integer spins.
Such mode is very crucial in the $AdS$/$CFT$ correspondence.

For fermions the situation is similar. Here, the spectrum
consists of an infinite tower
 of massive states and the question of  massless mode depends on the
 boundary conditions chosen. Indeed, there are boundary conditions,
 which do not mix 4D left and right-handed modes and lead to massless
 modes of definite chirality.

 Another issue we have discussed here, is the interaction problem of the HS fields.
 We know, from the analysis in Minkowski  background, that HS fields do not
 have minimal consistent interactions, not even with gravity. The reason is that gauge
 invariance, a crucial issue for the consistent propagation of HS gauge fields, is
 generally lost on curved backgrounds. This is due to the appearance of the
 Weyl tensor in the gauge transformed HS-field equations, which
 cannot be cancelled, even after imposing gravitational equations.
 In the latter case, the
 Weyl contribution leads to break
 down of the gauge invariance on a general curved background. As a consequence, tree-level unitarity
 is lost and the  HS interacting theory is ill defined.
 For the case of $AdS$ spacetime, which is maximally symmetric
 and conformally flat (vanishing Weyl tensor), gauge invariance can be
 maintained. However, gravitational perturbations of the $AdS$ background
 will remove conformal flatness and thus, HS field will continue to have inconsistent
 gravitational interactions on the $AdS$ background.
 Nevertheless, we should stress here that if additional gaugings are introduced, the propagation of HS fields
in curved backgrounds can be perfectly well-defined leading to their consistent interaction
with gravity \cite{Vass},\cite{Vasss},\cite{Sezgin}.

 Here, we extend
 the proposal of~\cite{Porrati} in the case of the $AdS$ spacetime.
In~\cite{Porrati}, tree-level unitarity is maintained by considering massive
HS fields and introducing non-minimal interactions with gravity.
These interactions  cancel hard gauge-breaking terms, although gauge invariance
is still softly broken due to an explicit mass term.
The theory then is lacking of any
hard breaking terms at linearized level, which could violate tree-level
unitarity and the latter
is restored up to the Planck scale.
 In the case of the $AdS$ background, the HS gauge fields are massless.
 However, $AdS$ space has a scale, proportional to the cosmological constant
 $\Lambda$. This scale
 is explicitly seen in the HS-field equations and has the form of a mass
 term, although the fields are in fact massless (as there are two propagating helicity modes). This scale allows
 the introduction of non-minimal interaction
 terms similarly to the Minkowski case which can indeed preserve tree-level unitarity. We explicitly showed this in
 the case of fermionic HS fields. The analysis for bosonic fields is similar,
 although more complicated. It should be noted that
 the non-minimal coupling is non-analytic in $\Lambda$ so that the flat-space
 limit cannot be taken. In particular it should also be noted that the no-go theorem of \cite{WW} that states
 the impossibility of consistent coupling between HS and spin $2$ particles, is circumvented by the introduction
 of non-minimal interactions \footnote{We thank the referee for pointing this out to us.}.

We have also discussed possible couplings between the HS fields and the matter living at the boundary branes.
Such couplings are induced from the non-minimal interaction of HS and the Weyl tensor. We know from the analysis
of the Einstein equations in spaces of codimension one branes and the  Israel-Darmois junction conditions,
that the discontinuities in the extrinsic curvature are proportional to the local matter distribution at the
points where discontinuities appear, i.e., at the brane positions~\cite{shiromitzu}.
As the Weyl tensor is written in terms of the curvature tensors, there is an  induced coupling
 between the HS fields and the
extrinsic curvature. Then, this coupling can be written as a local interaction term of the HS fields
and the matter living on the brane. Clearly, as the Weyl tensor is traceless, only the traceless part of the brane
energy-momentum tensor can be coupled to the HS fields, which is exactly what was found.

Finally, in the last part of this work, we have discussed the $AdS$/$CFT$ correspondence for HS fields. In this case, the HS bulk field equations
are solved with Dirichlet boundary condition and then the action is evaluated on the solution.  This procedure
gives the two-point function of HS operators of the (unknown) boundary $CFT$. We have followed this line for bosonic HS and we indeed
obtained the two-point function of boundary transverse traceless operators.
The same procedure in the RS background have been followed for
fermionic HS fields. In this case, the same procedure produces the two-point function for boundary operators of the
resulting $CFT$ on the boundary. In particular we have showed that the conformal invariance
of the boundary operators in the holographic $CFT$ at the boundary is restored once the IR brane
is removed.

As a final comment, we should mention the potential importance of HS fields in cosmology. In particular the fact that
HS fields interact with matter only gravitationally and not via gauge interactions make them possible dark-matter
candidates.

\vskip.5in

\noindent {\bf Acknowledgement} One of the authors (CG) would like
to thank F. Riccioni and P. J. Heslop for useful discussions. We would also like to thank A. Sagnotti for correspondence.
This work was
partially supported by the EPAN projects, Pythagoras and
Heraclitus, the NTUA programme Protagoras and the EC project
MRTN-CT-2004-005104. CG would like to
thank NTUA for its hospitality. CG is supported by PPARC research grant PPA/P/S/2002/00208.

\end{document}